\documentstyle[amstex,eqsecnum,aps,multicol,epsfig]{revtex}
\begin{document}
\title{Entangled graphs II:
Classical correlations in multi-qubit entangled systems}
\author{Martin Plesch$^{1}$ and Vladim\'{\i}r Bu\v{z}ek$^{1,2}$}
\address{${}^{1}$ Research Center for Quantum Information,
Slovak Academy of Sciences, 845 11 Bratislava, Slovakia
\\
${}^{2}$Department of Mathematical Physics, National University of Ireland,
Maynooth, Co. Kildare, Ireland
}
\date{April 4, 2003}
\maketitle

\begin{abstract}
Bipartite correlations in multi-qubit systems cannot be shared
freely. The presence of entanglement or classical correlation on
certain pairs of qubits may imply correlations on other pairs. We
present a method of characterization of bi-partite correlations in
multi-qubit systems using a concept of entangled graphs that has
been introduced in our earlier work [M.Plesch and V.Bu\v{z}ek,
{\em Phys. Rev. A} {\bf 67 }, 012322 (2003)]. In entangled
graphs each qubit is represented by a vertex while the
entanglement and classical correlations are represented by two
types of edges. We prove by construction that any entangled graph
with classical correlations can be represented by a {\em mixed}
state of $N$ qubits. However, not all entangled graphs with
classical correlations can be represented by a pure state.
\end{abstract}


\begin{multicols}{2}

\section{Introduction}

The laws  of quantum mechanics impose strict bounds on bi-partite
entanglement in multi-partite systems. This issue has been first addressed
by Wootters et al. \cite{Coffman2000,OConnor2001} who have derived
bounds on shared bi-partite entanglement in multi-qubit systems. In
particular, in their paper O'Connors and Wootters \cite{OConnor2001} have
searched for a state of a multi-qubit ring with maximal possible
entanglement between neighboring qubits. Another version of the same problem
has been analyzed by Koashi et al. \cite{Koashi2000} who have derived an
explicit expression for the multi-qubit completely symmetric state
(entangled web) in which all possible pairs of qubits are maximally
entangled.

A more general approach has been suggested by D\"{u}r who has
introduced a concept of {\em entanglement molecules}
\cite{Dur2000}, that is, quantum structures such that each qubit
is represented by a point (``atom'') while an  entanglement
between two qubits is represented by a  ``bound''. D\"{u}r has
shown that under the condition that the ``strength'' of the bound,
i.e. a particular value of  the degree of entanglement, is
arbitrary (though non-zero), an arbitrary entanglement molecule
can be represented by a {\em mixed} state of a multi-qubit system.
On the other hand, D\"{u}r has considered just the condition of
inseparability for a given set of pairs of qubits in the
multi-qubit system but he did not impose a strict condition of
separability for the remaining pairs of qubits. This issue has
been addressed in our earlier paper \cite{Plesch2002} where we
have introduced a concept of {\em entangled graphs}. In the graph,
each qubit is represented by a vertex and an edge between two
vertices denotes entanglement between these two qubits
(specifically, the corresponding two-qubit density operator is
inseparable). By construction we have proved that any entangled
graph with $N$ vertices and $k$ edges can be associated with a
{\em pure} multi-qubit state.

In
Refs.~\cite{Coffman2000,OConnor2001,Koashi2000,Dur2000,Plesch2002}
the main issue has been the distribution of bi-partite
entanglement in multi-qubit systems. On the other hand, it is of
importance to understand how entanglement as well as classical
correlations are shared in multi-qubits systems. In order to
illuminate this problem we generalize the concept of entangled
graphs. Specifically, we will consider entangled graphs with
classical correlations. In the graph each qubit is represented by
a vertex and vertices can be connected by two types of edges, one
type corresponds to entanglement between two specific qubits (the
corresponding bi-partite density operator is inseparable) while
the second type corresponds to classical correlations (the
corresponding bi-partite density operator is separable but not
factorized). The main result of our paper is that for any
entangled graph with classical correlations one can find a {\em
mixed} state that is represented by this graph. We also prove that
not every entangled graph with classical correlations can be
represented by a pure state; though we find several categories of
entangled graphs that can be associated with pure multi-qubit
states.

\section{Entangled graphs with classical correlations}
Let us consider a general state $\rho $ of an $N$-qubit system $S$. Density
matrices $\rho _{ij}$ of all possible pairs in the system $S$  are  defined as
\begin{equation}
\label{2.1}
\rho _{ij}={\rm Tr}_{S\backslash \{i,j\}}\left( \rho \right) ,
\end{equation}
where the trace is performed over the set of qubits
$S\backslash \{i,j\}$ which denotes the whole system except the two
qubits $i$ and $j$. In general, there exist two basic types of bi-partite
density matrices. Those fulfilling the separability condition (e.g.
see Refs.~\cite{Peres1996})
\begin{equation}
\rho _{ij}=\sum_{n}\zeta _{i}^{n}\otimes \xi _{j}^{n}
\label{2.2}
\end{equation}
are called separable, i.e., these density operators describe states
of two qubits that are not entangled but they are {\em classically}
 correlated (providing $n\geq 2$). All other states are entangled, i.e.
they are not separable.

In what follows our task will be to use the concept of entangled
graphs \cite{Plesch2002} to characterize bi-partite correlations
in  multi-qubit systems.  Firstly we note that when no
entanglement between two qubits is present then  two classes of
bi-partite density operators can be identified. These are: 1)
separable density operators of the form (\ref{2.2}), and 2)
density operators that are given by tensor products of single
particle density operators (i.e. the corresponding two-qubit
density operator is factorized \cite{Bouda2002}). Correspondingly,
we will  divide a set  of separable density matrices (\ref{2.2})
into
two categories: If the sum in (\ref{2.2}) has at least two terms ($%
n>1 $), the corresponding density operators describes classically
correlated bi-partite states. On the other hand if  there is only one
term present in the right-hand-side of equation (\ref{2.2}) then
\begin{equation}
\rho _{ij}=\rho _{i}\otimes \rho _{j},
\label{2.3}
\end{equation}
i.e., the bi-partite density operator is equal to the tensor
product of two single-qubit density operators and the
corresponding two qubits are not correlated at all. The
single-qubit density operators in Eq.~(\ref{2.3}) are obtained by
the standard trace rule
\begin{equation}
\label{2.4}
\rho _{i}=Tr_{S\backslash \{i\}}\left( \rho \right),
\end{equation}
where $S\backslash \{i\}$ denotes the set of all qubits except the $i$-th
one. If the condition (\ref{2.3}) is fulfilled then
the corresponding two qubits
are not correlated at all.

We remind ourselves that in the case of {\it entangled graphs }
\cite{Plesch2002} (where only entangled and separable matrices were
considered), an edge between two vertices has represented entanglement,
whereas no edge has simply meant no entanglement. In what follows we will
consider three types of bi-partite density operators: 1) non-separable
density operators that describe entangled pairs of qubits; 2) separable
density operators that describe classically correlated pairs of qubits, and
3) factorable density operators that describe states of independent
(uncorrelated) qubits. Therefore we will need two types of edges - one that
corresponds to the entanglement between two qubits, while the second type of
edge corresponds to separable qubits that cannot be described by factorized
density operators.

\begin{figure}[tbp]
\centerline {\epsfig{width=8.0cm,file=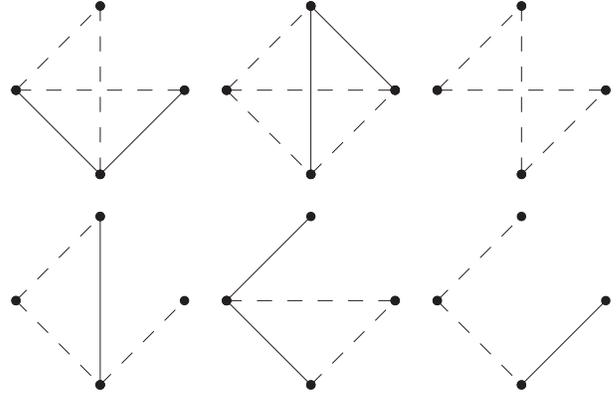}}
\medskip
\caption{Some examples  of entangled graphs with classical correlation
corresponding to states of four qubits. Solid edges are associated with
entangled bi-partite states, while dashed edges describe correlated but
not entangled (i.e. separable but not factorized) bi-partite states.
}
\label{fig1}
\end{figure}

With these two types of edges we can introduce the concept of an
entangled graph with classical correlations (some examples of such
graphs are presented in  Fig.~\ref{fig1}):
\begin{itemize}
\item  A system of $N$ qubits is represented by a graph with $N$ vertices;

\item  Vertices in the graph can be connected by two types of edges;

\item  Entanglement edge between two vertices (solid line) denotes non-zero
entanglement between relevant qubits; presence of entanglement implies also
the presence of classical correlation;

\item  Correlation edge between two vertices (dashed line) denotes classical
(and only classical) correlation between relevant qubits that are described
by separable but not factorable density operator [see Eq.~(\ref{2.2})];

\item  No edge between two vertices (no line) denotes no correlation between
relevant qubits and the corresponding density operator is given by the
tensor product of single-qubit density operators
[see
Eq.~(\ref{2.3})].
\end{itemize}

By definition, for a given multi-partite state (pure or mixed) it is always
possible to construct a corresponding graph. We simply calculate all
bi-partite density operators and test for the presence of entanglement, as
well as for the condition (\ref{2.3}) associated with the absence of
 classical correlation. However, the
inverse question is much more attractive: Given an {\it entangled graph with
classical correlations}, is it possible to construct a state, which would be
represented by that graph?  This question implicitly contains another
important issue: Does entanglement and classical correlations between
specific pairs of qubits imply entanglement and/or classical correlation on
other pairs of qubits in multi-qubits systems?

A graph corresponding to $N$ qubits
is completely specified by two sets of non-ordered pairs of
vertices $\{i,j\}$. The first set $S^{E}$ corresponds to entangled
pairs; $\{i,j\}\in S^{E}\Leftrightarrow \{i,j\}$ are entangled.
The second set $S^{C}$ describes correlated pairs; $\{i,j\}\in
S^{C}\Leftrightarrow \{i,j\}$ that are correlated. Each pair
$\{i,j\}\notin S^{C}$ is completely uncorrelated, i.e. it is in a
product state.
It is worth remembering that $S^{E}\subset S^{C}$, i.e. each entangled pair
is also classically correlated. We can define also a specific subset of
$S^{C}$, the set of classically (and only classically) correlated pairs
$S^{CC}=S^{C}\backslash S^{E}$. We define also a vector (of the length $N$)
$\overrightarrow{m}$, whose components $m_{i}$ denote the number of
qubits, that are uncorrelated with the $i$-th qubit; e.g., $m_{i}$ is
the number of pairs $\{i,j\}\notin S^{C}$ with fixed $i$. Let us denote
$M=\frac{1}{2}
\sum_{i=1}^{N}m_{i}$ as the total number of uncorrelated pairs.
The inequalities
\begin{eqnarray}
\nonumber
0 &\leq &m_{i}\leq N-1  \label{mi}\, ;  \\
0 &\leq &M\leq \frac{N(N-1)}{2}
\label{2.5}
\end{eqnarray}
exhibit simple attributes of the system, that no particle can be
uncorrelated with more  than $(N-1)$ particles,  and
that the maximum number of pairs of qubits in the system is
equal to $N(N-1)/2$.

\section{Mixed states}

A mixed state of a quantum mechanical system is always determined
by a larger number of parameters  than a pure state
of the same system. For instance, a pure state of a qubit is represented
by a point on a Poincare sphere, that is, each pure state is determined by
two parameters. On the other hand a mixed state (a convex combination of
pure states) is represented by a point inside a Poincare sphere and is
determined by three parameters. In general, number of parameters that are
needed for a specification of a mixed state is much larger than the number
of parameters needed for specification of a pure state.
One of the
consequences of this property of mixed states is that it is much
easier to fulfill constraints imposed by the graph structure on mixed states
with more ``free parameters'', than on pure states.

In what follows we present  a mixed state, which is defined by the sets
$S^{C}$ and $S^{E}$. Then we will prove that the bi-partite density operators
have all the desired properties, thus this state is  represented by
the graph specified by the sets $S^E$ and $S^C$.

\vspace{0.4cm}
{\em The mixed state of $N$ qubits given by the expression}
\vspace{-0.2cm}
\begin{align}
\rho & =\frac{1}{2\left( N-1\right) ^{2}}\{\left[ N^{2}-3N+\frac{1}{2}M+2%
\right] |0...0\rangle \langle 0...0|  \nonumber \\
& +\sum_{i=1}^{N}\left[ (N-1)-\frac{1}{2}m_{i}\right] |0...01_{i}0...0%
\rangle \langle 0...01_{i}0...0|  \nonumber \\
& +\sum_{\{i,j\}\in S^{E}}|0...01_{i}0...0\rangle \langle 0...01_{j}0...0|
\nonumber \\
& +\sum_{\{i,j\}\in S^{E}}|0...01_{j}0...0\rangle \langle 0...01_{i}0...0|
\label{3.1} \\
& +\sum_{\{i,j\}\notin S^{C}}\frac{1}{2}|0...01_{i}0...01_{j}0...0\rangle
\langle 0...01_{i}0...01_{j}0...0|\}  \nonumber
\vspace{-0.2cm}
\end{align}
{\em is characterized by a graph, specified  by the number of vertices $N$
and the sets $S^{E}$ and $S^{C}$.}
\vspace{0.4cm}

The density operator
(\ref{3.1}) is represented by
a convex sum of pure states $|0...0...0\rangle $,
$|0...01_{i}0...0\rangle $, $|0...01_{i}0...01_{j}0...0\rangle $ and
$\frac{1%
}{\sqrt{2}}\left( |0...01_{i}0...0\rangle +|0...01_{j}0...0\rangle
\right) $ so it describes a mixed state of $N$ qubits.

In what follows
we  show that for $\{i,j\}\in S^{E}$ the qubits $i$ and $j$ are
entangled. In this case, the reduced (bi-partite) density operator,
obtained from Eq.~(\ref{3.1}) by tracing over relevant qubits,
has the form
\begin{eqnarray}
\rho _{ij}^{E} &=&\frac{1}{2\left( N-1\right) ^{2}}
\label{3.2} \\
&&\times \left(
\begin{array}{cccc}
2N^{2}-6N+4 & 0 & 0 & 0 \\
0 & N-1 & 1 & 0 \\
0 & 1 & N-1 & 0 \\
0 & 0 & 0 & 0
\end{array}
\right) .  \nonumber
\end{eqnarray}
One could use the Peres-Horodecki criterion \cite{Peres1996} to
determine whether the density matrix (\ref{3.2}) describes an
entangled state of two qubits. Instead of this we can calculate
the concurrence \cite{foot1} of that state, that would allow us to
determine the strength of the entanglement. For the density matrix
(\ref{3.2}) the concurrence reads $\frac{1}{\left( N-1\right)
^{2}}$, thus it is larger than zero which means that the pair of
qubits $i$ and $j$ is indeed entangled.

For every $\{i,j\}\in S^{CC}=S^{C}\backslash S^{E}$ the qubits $i$ and $j$ have
to be correlated, but not entangled. We can calculate the corresponding
reduced density operator
\begin{eqnarray}
\rho _{ij}^{C} &=&\frac{1}{2\left( N-1\right) ^{2}}
\label{3.3} \\
&&\times \left(
\begin{array}{cccc}
2N^{2}-6N+4 & 0 & 0 & 0 \\
0 & N-1 & 0 & 0 \\
0 & 0 & N-1 & 0 \\
0 & 0 & 0 & 0
\end{array}
\right) .  \nonumber
\end{eqnarray}
This matrix is diagonal and the partial transposition would not
change it at all. This proves that the corresponding bi-partite
state is not entangled.

From Eq.~(\ref{3.3}) we can find
density operators $\rho_{i}$ and $\rho_j$ of individual  qubits:
\begin{equation}
\rho _{i}=\rho _{j}=\frac{1}{2\left( N-1\right) ^{2}}\left(
\begin{array}{cc}
2N^{2}-5N+3 & 0 \\
0 & N-1
\end{array}
\right) .  \label{3.4}
\end{equation}

In order to test the presence of
classical correlations we will utilize  the
condition (\ref {2.3}). The tensor product of  two states
(\ref{3.4}) corresponds to uncorrelated (factorized) two-qubit
density operator
\begin{eqnarray}
\rho _{i}\otimes \rho _{j} &=&\frac{1}{2\left( N-1\right) ^{2}}
\label{3.5} \\
&&\times \left(
\begin{array}{cccc}
2N^{2}-6N+\frac{9}{2} & 0 & 0 & 0 \\
0 & N-\frac{3}{2} & 0 & 0 \\
0 & 0 & N-\frac{3}{2} & 0 \\
0 & 0 & 0 & \frac{1}{2}
\end{array}
\right)  \nonumber
\end{eqnarray}
and we immediately see that $\rho _{i}\otimes \rho _{j}\neq \rho
_{ij}^{C}$, i.e. the pair of qubits in the state (\ref{3.3})  is
correlated.

For the rest of the pairs $\{i,j\}\notin S^{C}$ the reduced density operator
$\rho_S$
can be found to be given by Eq.~(\ref{3.5}) which means that these qubits
are not correlated at all since
$\rho _{ij}^{S}=\rho_{i}\otimes \rho _{j}$.

Herewith we have proved that the state (\ref{3.1}) is represented
by a graph specified by the two sets $S^{E}$ and $S^{C}$. The
state (\ref{3.1}) exhibits also some
other interesting properties. For instance, concurrencies  for all pairs $%
\{i,j\}\in S^{E}$ have the same value. This is a natural
consequence of the fact that all the density matrices of entangled
pairs are identical. From Eq.~(\ref {3.4}) we also see that all
single-qubit density operators are identical and that they depend
only on the total number of qubits $N$. This means that the
information about the graph itself is encoded only in the
correlations; there is no way to extract any information about the
specification of the graph only via single-qubit measurements.

\section{Pure states}

The problem of a construction of {\em pure} states corresponding
to a specific graph is (as expected) more complicated than for
mixtures. As mentioned above, 
the number of ``free'' parameters in this case is
smaller  and one cannot ``control''
off-diagonal matrix elements in the same way as in the case of
mixed states \cite{foot2}. Therefore, we start our discussion with
the simplest case of three qubits and we examine thoroughly all
possible graphs. Then we formulate a theorem about the existence
and non-existence of some classes of graphs.

\subsection{Three-qubit graphs}

\begin{figure}[tbph]
\centerline {\epsfig{width=8.0cm,file=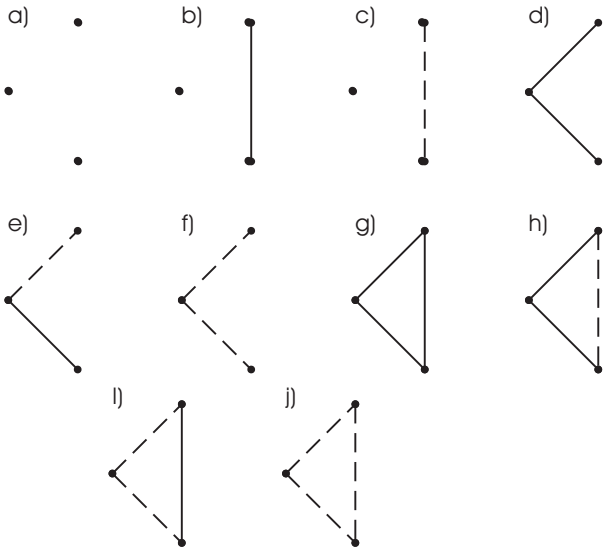}}
\medskip
\caption{Ten possible entangled  graphs with classical correlations
for three qubits. Graphs c), d), e), f) do not have
representatives among pure states.}
\label{fig2}
\end{figure}

In the case of three qubits
there are ten possible entangled graphs with classical correlation.
We present these graphs in Fig.~\ref{fig2}.
We know, that for six of these graphs there exist pure states. For example
\begin{eqnarray*}
{\rm a)}
&\rightarrow &|000\rangle \, ;\\
{\rm b)}
&\rightarrow & \frac{1}{\sqrt{2}}|0\rangle(|00\rangle+|11\rangle)\, ; \\
{\rm g)}
&\rightarrow &\frac{1}{\sqrt{3}}(|001\rangle+|010\rangle+|100\rangle)\, ; \\
{\rm h)}
&\rightarrow &\frac{1}{2}\left( |000\rangle +|100\rangle +|110\rangle
+|111\rangle \right)\, ; \\
{\rm i)}
&\rightarrow &\frac{1}{\sqrt{3}}\left( |000\rangle +|011\rangle
+|111\rangle \right)\, ; \\
{\rm j)}
&\rightarrow &\frac{1}{\sqrt{2}}(|000\rangle+|111\rangle).
\end{eqnarray*}
For the other four graphs it is not possible to find any pure
state, which would be represented by them. These four graphs
exhibit one common property: they all include at least one vertex,
which is connected with just one of the other two vertices. We
will show, that this property of a graph immediately leads to
non-existence of a pure state associated with this graph. The only
possible exception is a graph of the form b) in Fig.~\ref{fig2},
when there is an isolated pure entangled two-qubit state (as for
example a Bell pair $\frac{1}{\sqrt{2}}(|00\rangle+|11\rangle)$),
which is not connected with the rest of the system at all.

\subsection{Multi-qubit graphs}

For more than three qubits we have a very large number of possible
graphs.  Typically the number of different graphs grows as
$\exp(N^{2})$ and already for $N=4$ it gives a number greater than
$100$. Necessarily, one needs to categorize these graphs in order
to study the problem. Therefore, let us divide these graphs into
two basic categories:

\begin{itemize}
\item  {\em Disconnected graphs}: These are the graphs whose
vertices can be divided into (at least) two groups (each group
containing at least one vertex), which are connected neither by an
entanglement edge, nor by a correlation edge.

\item  {\em Connected graphs}:
In these graphs every pair of vertices is connected
directly or via other vertices.
\end{itemize}

Let us first consider  disconnected graphs. For a large $N$ this
group of graphs is much smaller than the second one. For
disconnected graphs the question of existence of pure states can
be easily reduced to a problem of graphs with smaller number of
vertices.

\begin{figure}[tbp]
\centerline {\epsfig{width=6.0cm,file=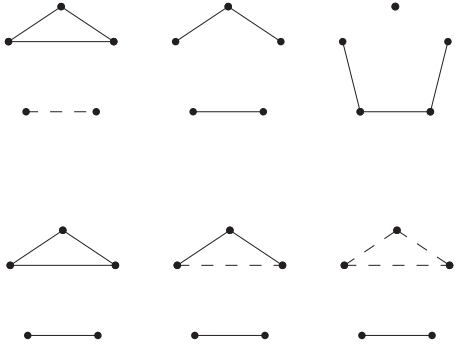}}
\medskip
\caption{Six examples of disconnected entangled graphs with
classical correlations for five qubits. In the first row there are
examples of graphs that  cannot be represented by
 pure states of five qubits. In the second row there are
examples of graphs that can be represented by pure states. }
\label{fig3}
\end{figure}

We can divide all  vertices in a disconnected graph into two
groups, which are not connected by any edge. We denote the two
subsystems $A$ and $B$, respectively. As the two subsystems are
not correlated, we can write
\begin{eqnarray}
\label{4.1}
\rho _{AB}=\rho_{A}\otimes \rho_{B},
\end{eqnarray}
where $\rho _{AB}$ is the density operator of the whole system while
$\rho
_{A}$ and $\rho _{B}$ are the density operators of the two subsystems.
According to our assumption the whole system is in a pure state, i.e. $%
\rho _{AB}$ is pure. Consequently,  the two subsystems have to be in pure
states as well. Thus
\begin{equation}
\left| \Psi \right\rangle _{AB}=\left| \psi \right\rangle _{A}\otimes \left|
\psi \right\rangle _{B}
\label{4.2}
\end{equation}
and we can state that the whole state $\left| \Psi \right\rangle
_{AB}$ exists, iff $\left| \psi \right\rangle _{A}$ and $\left|
\psi \right\rangle _{B}$ do exist. One could follow the same
argument if  there are more disconnections in the graph.
 Therefore, every disconnected graph can be
represented by a pure state if and only if every separated subset
of vertices (separated subgraph) can be represented by a pure
state. In Fig.~\ref{fig3} we present
examples of  disconnected graphs that cannot be represented by pure
states (first row) and that can be represented by pure states (second row).

For connected graphs we have not been able to find any simple
algorithm to determine  the existence of a pure state that  would
represent a given graph. However, we can formulate theorems about
specific classes of graphs, which exhibit some special properties.

It is obvious, that every graph containing only one vertex has a
representation among pure states (any pure state of a qubit). Also
for a graph containing two vertices, that are connected by an
entanglement edge one is able to find a pure state (for example a
Bell-state $\frac{1}{\sqrt{2}}(|00\rangle+|11\rangle)$). On the
other hand, there is no pure state that would correspond to a two-vertex
graph with a correlation edge
This can be seen from the
fact, that such a state would have to be written in a form
\begin{equation}
\left | \psi \right \rangle _{AB} \left \langle \psi \right|=
\sum_{i=1}^{k} \lambda_{i} \left (\rho_{A}^{i} \otimes
\rho_{B}^{i} \right),
\end{equation}
with $k>1$, $\lambda_i>0$ and $\rho_A^i\neq\rho_A^j$, what is
clearly not possible.

For more qubits, we can determine one type of graph that 
cannot be represented by pure states: These are the graphs with
the so-called {\it open edges}. 
If a vertex in a connected multi-vertex
graph  is connected with
the rest of the graph with just a single edge (correlation or
entanglement), then we will call it as an {\it open edge}  (since
it is not a part of any closed chain of edges). 

Now we can
formulate a theorem for graphs with open edges:

\vspace{0.4cm}
{\em A connected graph with $N$ vertices ($N>2$) containing at least one
open edge can never be represented by a pure state of $N$ qubits.}
\vspace{0.4cm}

\begin{figure}[tbp]
\centerline {\epsfig{width=4.0cm,file=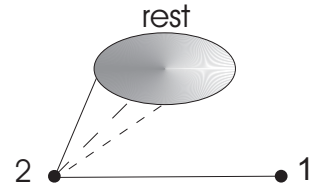}}
\medskip
\caption{Schematic visualization of the structure of a graph with
an open edge. The vertex 1 is connected with the rest of the
system via a single edge.} \label{fig4}
\end{figure}

To prove this theorem  let us denote the qubit, which is connected
only via one edge (the open edge) with the rest of the system, as the first
qubit. The qubit mediating this connection (the other end of
the open edge) will be denoted as the second one. The structure of the
remaining part of the graph is not important for our consideration and
we simply denote it as the ``rest'' (see Fig.~\ref{fig4}).

Since we consider the whole $N$-qubit system to be in a pure state
the state of the first qubit has to be mixed (otherwise it could not be
correlated or entangled with other parts of the system). The density
operator of the first qubit then can be written in a form
\begin{equation}
\label{4.3}
\rho _{1}=a\left| \psi \right\rangle \left\langle \psi \right| +\left(
1-a\right) \left| \psi^\perp \right\rangle \left\langle \psi^\perp \right| ,
\end{equation}
where $0<a<1$ and the two states $|\psi\rangle$ and
 $|\psi^\perp\rangle$ are mutually orthogonal, i.e.
$\langle\psi|\psi^\perp\rangle=0$.

The part of the graph  denoted as the ``rest'' with  $N-2$ vertices
is also in a mixed state with the corresponding density operator
$\rho_{rest}$ that can be written in a form
\begin{equation}
\rho _{rest}=A\left| \Psi \right\rangle \left\langle \Psi \right|
+(1-A)\rho_{\Psi}^\perp   \, ,
\label{4.4}
\end{equation}
where $0<A<1$ and $\rho_{\Psi}^\perp$ is a density operator of $N-2$ qubits
which is orthogonal to the state $|\Psi\rangle$, i.e.
$\langle\Psi|\rho_{\Psi}^\perp|\Psi\rangle=0$.
Because we assume that the part of the graph (corresponding to $N-2$ qubits)
that we denote as ``rest'' is not correlated with the vertex 1 at all,
we can express the joint density operator $\rho _{1\oplus rest}$
of the first qubit and the part ``rest'' as the tensor product of two
density operators $\rho _{1}$ and $\rho _{rest}$, i.e.
\begin{eqnarray}
\rho _{1\oplus rest}= \rho _{1}\otimes \rho _{rest}\, .
\label{4.5}
\end{eqnarray}

On the other hand, by the definition of our task  the whole
graph corresponding to $N$ qubits has to be in a pure state
$|\Xi\rangle_{1\oplus2\oplus rest}$.
Correspondingly, the qubit (vertex) number ``2'' has to purify
simultaneously both density operators $\rho_1$ and $\rho_{rest}$
in such a way that $\rho_{1}={\rm Tr}_{2,rest} (|\Xi\rangle\langle\Xi|)$
and $\rho_{rest}={\rm Tr}_{2,1} (|\Xi\rangle\langle\Xi|)$, while
$\rho _{1\oplus rest}={\rm Tr}_2 (|\Xi\rangle\langle\Xi|)$.

However, this is impossible even if we assume that the density
operator $\rho^\perp_\Psi$ in Eq.~(\ref{4.4}) is a projector (i.e.
$\rho^\perp_\Psi=|\Psi^\perp\rangle \langle\Psi^\perp|$) since
even in this case the density operator $\rho _{1\oplus rest}$ is
equal to a statistical mixture of four mutually orthogonal states:
\begin{eqnarray}
\rho _{1\oplus rest}&=&\rho _{1}\otimes \rho _{rest}
\label{4.6} \\
&=&aA\left| \psi \Psi \rangle \langle \psi \Psi \right|
+a(1-A)\left| \psi \Psi^\perp \rangle \langle \psi \Psi^\perp \right|
\nonumber \\
&& +(1-a)A\left| \psi^\perp \Psi\rangle \langle \psi^\perp \Psi \right|
\nonumber \\
&& +
(1-a)(1-A)\left| \psi^\perp \Psi^\perp \rangle \langle \psi^\perp\Psi^\perp
\right|\, .
\nonumber
\end{eqnarray}
As discussed earlier in Sec.~III
in order to purify the state (\ref{4.6}) we would need a four-dimensional
ancilla \cite{Uhlmann1976}, which obviously is not available in our
considerations since the vertex ``2'' is just a qubit with a 2-dimensional
Hilbert space. This proves the Theorem 2.

\subsection{Other classes of graphs}

{\em Entangled webs}\newline Let as consider graphs with {\em all}
pairs of vertices connected with an edge (either correlation or
entanglement). These types of graphs can be represented by pure
states of the form
\begin{eqnarray}
\label{4.7}
|\Xi \rangle &=&
\alpha |0..0\rangle +\beta |1..1\rangle
\\
&&
\nonumber
+\sum_{\{i,j\}\in S^E}%
\frac{\gamma }{\sqrt{k}}|1\rangle _{i}|1\rangle _{j}
|0...0\rangle_{S\backslash \{i,j\}}
\end{eqnarray}
with the normalization condition
$\left| \alpha \right| ^{2}+\left| \beta \right|
^{2}+\left| \gamma \right| ^{2}=1$, and $\alpha, \beta, \gamma > 0$.
 The pure state (\ref{4.7}) describes
a graph such that  pairs of vertices in the set $S^E$ are
entangled while all other pairs of vertices are correlated
\cite{Plesch2002}.

\medskip

{\em Connected graphs with no open edges}
\newline
Through numerical simulations we have searched for pure states
corresponding to graphs of four qubits. We have found a pure state
for every connected graph with no open edges. Specifically, taking
into account general properties of graphs described above we have
found only $20$ graphs that remain ambiguous. With the help of our
simulation we have been able to find at least one pure state as a
representative of each  graph.

Unfortunately, we have not been able to generalize this result for
connected graphs with no open edges for more than four vertices.

\section{Conclusions}

In order to understand how correlations and entanglement are shared among
qubits in multi-qubit systems we have introduced a concept of entangled
graphs with classical correlations.
Every qubit is represented by a vertex and
correlations between two qubits are represented by  edges. Two
types of edges stand for two possible types of (non zero)
correlations: the entanglement edge corresponds to
 entanglement between a specific pair of qubits (vertices), while
the correlation edge denotes classical correlation. No edge
between two qubits means that the corresponding bi-partite density
operator is the tensor product of single-qubit density operators.

We have shown that  any graph with $N$ vertices can be represented
by  a mixed state of $N$ qubits. On the other hand only some
graphs can be represented by pure states. In particular, we have
shown that  connected graphs with $N$ vertices that contain an
open edge can never be represented by a pure state. Interestingly
enough, in the case of three- and four-vertex graphs we have been
able to find pure states for all other graphs (i.e. connected
graphs with no open edges).

\acknowledgements
This work was supported by the European Union projects QUPRODIS
and QGATES. VB would like to acknowledge a support from 
the Science Foundation Ireland.

\end{multicols}


\begin{references}
\bibitem{foot2}
The
off-diagonal elements are the biggest ``problem'' for classical
correlation. The reason is  that if one type of measurements (let's say
$\sigma _{x}$) is not correlated, the other type (such as $\sigma _{z}$) can
be.
\end{references}
\end{document}